\documentclass[english]{article}
\usepackage[T1]{fontenc}
\usepackage[latin9]{inputenc}
\usepackage{textcomp}
\usepackage{amstext}
\usepackage{amsmath}
\usepackage{graphicx}
\makeatletter
\usepackage[auth-lg]{authblk}
\usepackage{babel}
\usepackage{cite}
\usepackage{epstopdf}

\usepackage{newfloat}
\DeclareFloatingEnvironment[name={Supplementary Figure}]{suppfigure}

\usepackage[margin=1in]{geometry}

\begin{document}

\title{Programming filamentous network mechanics by compression}

\author[a]{Bart E. Vos}
\author[b]{Luka C. Liebrand}
\author[b]{Mahsa Vahabi}
\author[b]{Andreas Biebricher}
\author[b]{Gijs J. L. Wuite}
\author[b]{Erwin J. G. Peterman}
\author[a,c]{Nicholas A. Kurniawan}
\author[b,d,e]{Fred C. MacKintosh\footnote{To whom correspondence should be addressed. E-mail: fcmack@rice.edu}}
\author[a]{Gijsje H. Koenderink\footnote{To whom correspondence should be addressed. E-mail: g.koenderink@amolf.nl}}

\affil[a]{Biological Soft Matter group, FOM Institute AMOLF, Amsterdam, The Netherlands}
\affil[b]{Department of Physics and Astronomy, Vrije Universiteit, Amsterdam, The Netherlands}
\affil[c]{Department of Biomedical Engineering \& Institute for Complex Molecular Systems, Eindhoven University of Technology, Eindhoven, The Netherlands}
\affil[d]{Departments of Chemical \& Biomolecular Engineering, Chemistry and Physics \& Astronomy, Rice University, Houston, TX 77005, USA}
\affil[e]{Center for Theoretical Biophysics, Rice University, Houston, TX 77030, USA}

\date{}

\maketitle


\textbf{Fibrous networks are ideal functional materials since they provide mechanical rigidity at low weight. Such structures are omnipresent in natural biomaterials from cells to tissues, as well as in man-made materials from polymeric composites to paper and textiles. 
Here, we demonstrate that fibrous networks of the blood clotting protein fibrin undergo a strong and irreversible increase in their mechanical rigidity in response to compression. This rigidification can be precisely predetermined from the level of applied compressive strain, providing a means to program the network rigidity without having to change its composition.
To identify the mechanism underlying this programmable rigidification, we measure single fiber-fiber interactions using optical tweezers. We further develop a minimal computational model of adhesive fiber networks that shows that load-induced bond formation can explain the adaptation of the fibrin networks to compressive loading.
The model predicts that the network stiffness after compressive programming obeys a universal power-law dependence on the prestress built in by new bond formation, which we confirm experimentally.
The generality of this functional stiffening mechanism together with our ability to quantitatively predict it provides a new powerful approach to program the stiffness of fibrous materials and dynamically adapt them to different loading conditions.}

Fibrous materials are ubiquitous in biology, forming the structural framework of cells and connective tissues \cite{Pritchard2014a}.
Similar design principles are also harnessed in many man-made materials ranging from paper and textiles to light-weight composites \cite{Picu2011}.
From a physical perspective, fibrous materials utilize a smart principle, since stiff polymers and rigid fibers can form space-filling elastic networks at exceedingly low volume fractions of less than 1\%, and at low connectivities where only 3 to 4 fiber segments meet at each node. This connectivity is below the Maxwell isostatic threshold for simple spring networks, which requires that an average of at least 6 springs should meet at nodes for a mechanically stable network. Fibrous materials can beat this threshold, e.g., due to the stabilizing effects of stress or the bending resistance of fibers \cite{Broedersz2011,Sharma2015a,Licup2015}. Hence, the stiffness of fibrous networks is inherently strongly sensitive to the density of cross-linkers between fibers \cite{Gardel2004}. Furthermore, fibrous networks reversibly stiffen under shear deformation, a phenomenon that has been explained by network models consisting of either semiflexible or elastic fibers \cite{Statics2005,Licup2015,Sharma2015a}.
These properties in principle allow one to adjust the stiffness of fibrous networks to specific functional requirements. However, they offer only limited options for dynamic adjustment of the mechanical rigidity of fibrous materials to diverse functional requirements since changing cross-link density requires making a new system for each application, while strain-stiffening is only operative as long as a mechanical strain is applied.

Nature has found intriguing ways to adapt the mechanical performance of fibrous networks in tissues in a more dynamic fashion to diverse and time-varying mechanical loading conditions.
Tissues are able to actively reinforce their structure along the principal load direction. The mechanisms of mechanical reinforcement are generally thought to originate from cellular activity, involving strain-dependent fiber degradation and synthesis \cite{Dittmore2016,Araujo2011}. 
However, recent studies suggest that biopolymer networks are inherently adaptive themselves, since they are held together by weak transient bonds \cite{Gralka2015,Kurniawan2016}. Cyclic shear loading has been shown to cause reinforcement for a number of biopolymer systems \cite{Knight2006,Schmoller2010,Lopez-Menendez2016,Fernandez2008}, although softening can occur as well \cite{Munster2013a,Nam2016}. The physical principles responsible for these varied inelastic responses are still not fully understood, involving bond breaking and reformation between fibers \cite{Wolff2010,Nam2016,Kurniawan2016}, but also within fibers \cite{Munster2013a}.

\begin{figure*}[ht]
	\begin{center}
		\centerline{\includegraphics[width=.4\textwidth]{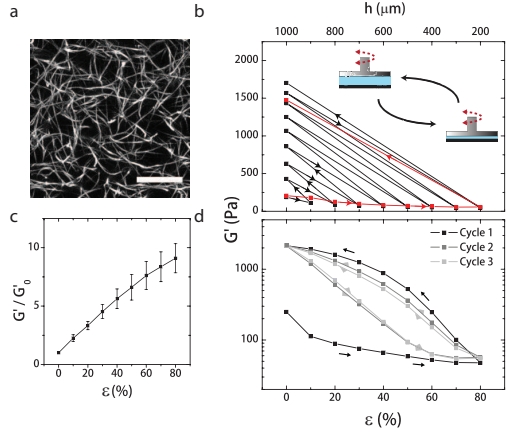}}
		\caption{\textbf{Cyclic compression experiments on fibrin networks.} 
			\textbf{a}, Confocal microscopy image of a fibrin network. (Maximum intensity projection of a stack of confocal slices acquired over a depth of 10~$\mu$m; scale bar, 10~$\mu$m).
			\textbf{b}, Shear modulus $G'$ of a fibrin network as a function of compressive strain $\epsilon$ and corresponding sample height $h$. The arrows indicate the directions of stepwise (de)compression. The inset show how the shear modulus is obtained by applying a small oscillatory shear strain while varying the axial strain. The \textbf{black} curve is obtained by cyclic loading-unloading in strain steps of 100 $\mu$m, while the \textbf{red} curve was obtained by directly compressing to $\epsilon = 80\%$. (De)compression was done slowly and with a waiting time after each step to ensure free water flow. We observe reversible strain-softening with increasing compression and irreversible stiffening with decompression. 
			\textbf{c}, Normalized network stiffening $G'/G'_0$ as a function of the maximum compressive strain $\epsilon$ that the network has experienced.
			\textbf{d}, Repeated compression to 80\% strain and decompression, showing the first (\textbf{black}), second (\textbf{grey}) and third (\textbf{light grey}) cycle. Arrows indicate the sequence of the compression-decompression steps.
		}\label{fig:Figure1}
	\end{center}
\end{figure*}

To test the influence of cyclic loading on the inelastic behavior of fibrous networks, we choose fibrin as a model biopolymer (Fig. \ref{fig:Figure1}a) because its elastic properties have already been extensively studied in the context of blood clotting \cite{Mosesson2005}. Here we show that fibrous networks composed of fibrin undergo strong, programmable stiffening in response to cyclic compression.
By directly measuring the interactions between individual fibrin fibers with optical tweezers, we reveal that this stiffening originates from load-induced formation of new bonds. We develop a minimal computational model of adhesive fiber networks to show that bond remodeling can indeed explain reinforcement of fibrin networks under compression. Furthermore, the model predicts that network stiffening obeys a general power-law dependence on the internal prestress that we controllably build up via bond formation during compressive loading. The generality of this training capability of adhesive fiber networks can be used as a novel basis for dynamically programming the stiffness of fibrous materials.


We polymerize fibrin networks between the plates of a shear rheometer and compress each network stepwise in increments of 10\% axial strain and decompress back to the original state (0\% axial strain) after each compression step. We perform each (de)compression step at a slow rate (1~$\mu m$~$s^{-1}$) to allow for water efflux and influx. Furthermore, we allow the network to equilibrate for at least 125 seconds until the normal force exerted by the network on the rheometer top plate reaches a constant level (Supplementary Fig. \ref{fig:SIFigure6}). We probe the rigidity of the network at different levels of axial strain by measuring the shear modulus with a small amplitude oscillation (shear strain amplitude $\gamma=0.005$ and oscillation frequency $\nu=0.5$ Hz). As we compress the network to increasing axial strains, the shear modulus progressively decreases (Fig. \ref{fig:Figure1}b, black). This softening response is consistent with previous studies of fibrin \cite{Kim2014,Rosakis2015,Oosten2016} and actin \cite{Chaudhuri2007} and may indicate fiber buckling
\cite{Landau1986,John2013,Rosakis2015}. Even for pure spring networks, extension tends to stabilize networks while compression tends to destabilize networks \cite{Alexander1998,Sheinman2012a}.
At the highest maximum compressive strains that we apply (80\%), there is a slight upturn of the modulus, reflecting network densification \cite{Kim2014,John2013}. Strikingly, when we decompress the network to its original height, we observe a strong increase in the shear modulus compared to the virgin state. The increase is already more than twofold when the network has experienced a compressive strain of 10\% and rises to eightfold when the network has experienced a compressive strain of 80\% (Fig. \ref{fig:Figure1}c). Importantly, the shear modulus reached after decompression is constant over time, so the compressive programming is irreversible. To test whether the loading history affects the inelastic response, we also programmed networks in a single cycle of compression and decompression (Fig. \ref{fig:Figure1}b, red). The softening upon compression and stiffening upon decompression are near-identical to the changes observed in the stepwise programming protocol, suggesting that only the maximum compressive strain that the sample has experienced is relevant, while the subsequent strain history plays no role. This conclusion is further supported by repeating cycles of compression to 80\% strain and decompression to 0\% strain, where we observe that the once-programmed network switches between the same high stiffness at 0\% strain  and low stiffness at 80\% strain in subsequent cycles (Fig. \ref{fig:Figure1}d).

\begin{figure*}[ht]
	\begin{center}
		\centerline{\includegraphics[width=.7\textwidth]{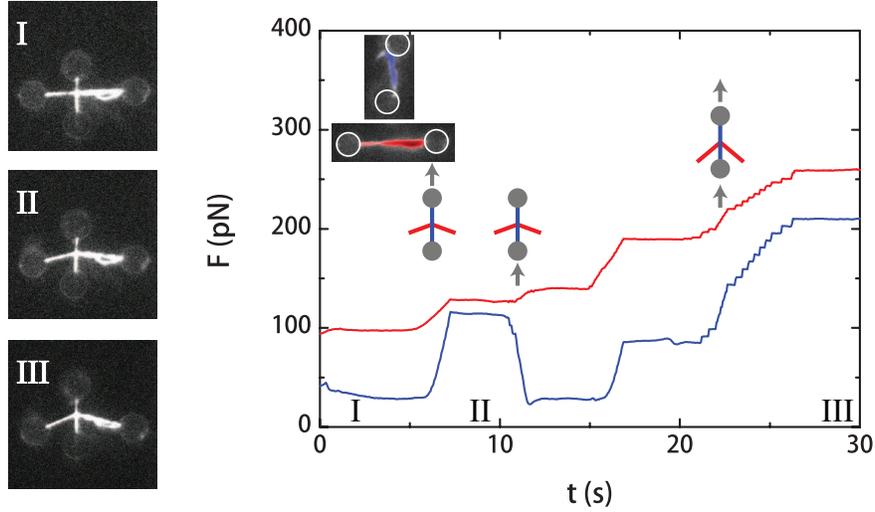}}
		\caption{\textbf{Direct measurement of the interaction between two individual fibrin fibers by optical tweezers.} Fluorescence microscopy images (left) and force ($F$) versus time ($t$) graph (right) for a typical optical tweezers experiment. The numbers above the fluorescence images correspond to time points indicated in the graph. Two fibers are crossed and brought into contact using four optically trapped beads as handles. After the vertically oriented fiber (\textbf{blue}) is brought in contact with the horizontally oriented fiber (\textbf{red}) (I), there is a force response on both fibers when the upper or lower bead (as illustrated by the sketch) of the vertical bead pair is moved. A movie showing the entire time-lapse sequence can be found in the \textit{Supplementary Information}. Error bars represent standard deviations.
		}
		\label{fig:Figure2}
	\end{center}
\end{figure*}

The strong irreversible stiffening of the fibrin networks upon axial compression suggests that the structure of the network is being remodeled. A plausible hypothesis is that the fibers form additional bonds during compression, since compression increases the fiber density and thereby enhances the chance of fiber-fiber interactions. To test this hypothesis, we develop a quadruple optical tweezers assay to directly probe the interaction between two individual fibrin fibers \cite{Laurens2012,VanMameren2009,Kurniawan2016}. Using a microfluidic flow cell with separate inlets for a dilute suspension of 4.5~$\mu$m-sized polystyrene beads, a dilute solution of fluorescently labeled fibrin fibers, and assay buffer, we first capture four beads in the traps, then suspend a fibrin fiber between each bead pair, and move the two fibers into the buffer channel. We next bring the fibers in a crossed configuration by rotating one fiber (vertical in the fluorescence images in Fig. \ref{fig:Figure2}), bringing it underneath the second, horizontally oriented fiber, and finally moving it upwards into contact with the horizontal fiber. To test whether the fibers spontaneously bond, we pull on the vertical fiber. As shown in Fig. \ref{fig:Figure2} and the corresponding Supplementary Movie 1, pulling on the vertical fiber causes displacement of the horizontal fiber. This observation provides clear evidence that the fibers spontaneously form bonds when brought in contact. To measure the bond strength, we monitor the force on the trapped beads. The red curve in Fig. \ref{fig:Figure2} shows the force as a function of time on the beads connecting to the horizontal, stationary fiber, while the blue curve corresponds to the vertical fiber. Each time we pull on the vertical fiber, there is a corresponding force increase on the horizontal fiber until, at a force of 347~pN, the uppermost bead is pulled out of the trap. Indeed, we observe consistently (10 independent experiments) that the bead is pulled out of the trap while the connection between the fibers remains unbroken. As the trap strength is in the range of 300-400~pN, the tweezers measurements set a lower limit on the strength of the newly formed fiber-fiber junction of $343\pm104$~pN.

What is the molecular mechanism of new bond formation? Fibrin fibers are thick bundles of around 65 double-stranded protofibrils held together by covalent and noncovalent interactions \cite{Kurniawan2016}. We polymerize fibrin in the presence of FXIII, a physiologically important enzyme that stabilizes blood clots by generating covalent peptide bonds between fibrin monomers. A possible origin of the spontaneous fiber bonding could be the creation of new cross-links by fibrin-bound FXIII as the fibers are brought in contact. To test this idea, we repeated the fiber interaction measurements in the presence of the specific FXIII inhibitor D004 \cite{Kurniawan2014}. We again observed strong fiber-fiber adhesion (Supplementary Fig. \ref{fig:SIFigure1}) and consistent with this, we observed the same extent of network stiffening after compressive loading as in FXIII-cross-linked networks (Supplementary Fig. \ref{fig:SIFigure2}). We conclude that FXIII induced cross-linking is not responsible for bond formation. Instead, we propose that bond formation is facilitated by noncovalent interactions, most likely mediated by the unstructured $\alpha$C-regions, two long and flexible chains attached to the distal ends of each fibrin monomer \cite{Protopopova2015}. Specific interactions between these chains mediate lateral association of protofibrils into fibers as well as interactions between fibrin fibers \cite{Litvinov2007}. We estimate that the total binding force between two interacting fibers mediated by adjacent $\alpha$C-regions is approximately $760$~pN, sufficient to account for the absence of junction rupture in the optical tweezers experiments, and the irreversibility of bond formation in the network-scale experiments (see Supplementary Information for the detailed calculation).



\begin{figure*}[ht]
	\begin{center}
		\centerline{\includegraphics[width=.7\textwidth]{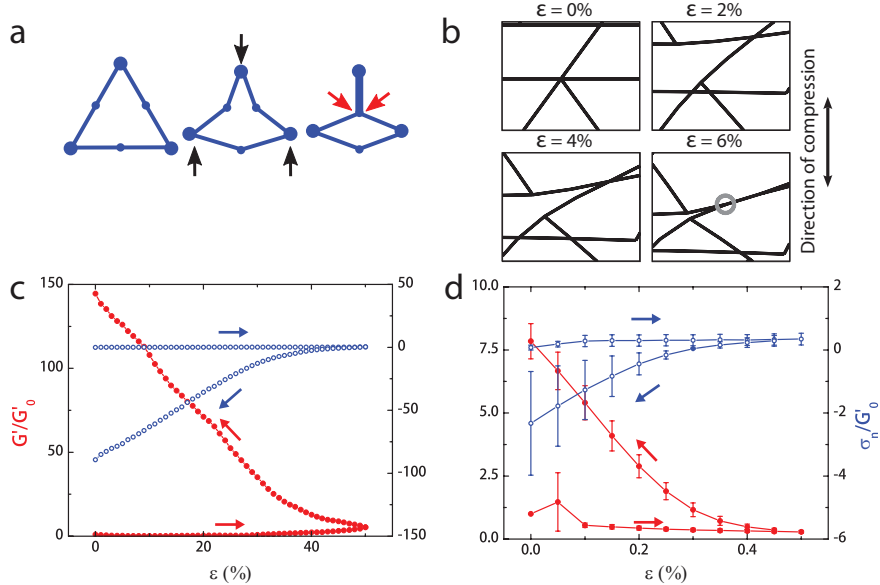}}
		\caption{\textbf{Minimal computational model of adhesive fibrous networks.} \textbf{a}, Fibrous networks are modeled as 2D disordered triangular lattices where every fiber contains one empty midpoint that can merge with another empty midpoint during compressive deformation to form a new junction. Under compression (black ), two nodes within a remodeling distance $d$ merge (red arrows). \textbf{b}, A small section of a simulated network during compression, starting from an initial state with no applied axial strain to the moment where the first merging event occurs (encircled in grey). A movie showing a complete compression sequence can be found in the \textit{Supplementary Information}. \textbf{c}, Shear modulus (\textbf{red}) and corresponding normal force (\textbf{blue}) as a function of compressive strain  for a simulated network ($\tilde{\kappa} = 10^{-4}$) and \textbf{d}, a fibrin network. The modulus and normal stress are both normalized by the initial shear modulus of the virgin network. Arrows indicate the sequence of the compression-decompression steps. Error bars represent standard deviations.
		}
		\label{fig:Figure3}
	\end{center}
\end{figure*}

The observation that fibrin fibers have the intrinsic ability to form strong bonds raises the question whether this mechanism is able to explain fibrin network reinforcement by compressive loading. To address this question, we develop a computational network model that allows for new bond formation under deformation. Fibrous networks are modeled as a 2D triangular disordered lattice with lattice spacing $l_0$ and dimensions of $50 l_0 \times 50 l_0$. The average connectivity is adjusted to 3.4 by phantomization and dilution, to reflect the combination of branches and cross-links that connect fibers in fibrin and other biopolymer networks (see Methods). Similar to the experiments, the lattices are compressed in steps of $1\%$ from an initial axial strain of $0\%$ to $50\%$ and back to $0\%$. The fibers are modeled as elastic beams with a stretching modulus $\mu$ and bending modulus $\kappa$. We performed simulations for a fixed fiber rigidity, $\tilde{\kappa}=\kappa / \mu l_{0}^{2} = 10^{-4}$, which lies in the relevant range for fibrin networks. After each (de)compression step, we minimize the energy of the network and determine the normal and shear stresses from the variation of the energy with strain. Periodic, Lees-Edward boundary conditions are used to minimize boundary effects. To allow for network remodeling, we introduce a single midpoint between adjacent network nodes and allow two midpoints to merge into one node when applied deformations force them to approach one another to within a predetermined capture radius $d$, which we will refer to as the remodeling distance. In our simulations we choose $d=0.01 l_0$. Fig. \ref{fig:Figure3}a illustrates such a merging event in half a unit cell of the lattice.

A typical remodeling event during the first steps of compression is highlighted in Fig. \ref{fig:Figure3}b, which shows a zoomed-in section of a simulated network. A non-trivial combination of bending and stretching of individual fibers brings two midpoints to within the remodeling distance, encircled in gray. Supplementary Movie 2 shows how the network as a whole responds to the applied compression.
With increasing compressive strain, the simulations predict an initial softening, followed by stiffening when the axial strain reaches ca. 30\% (Fig. \ref{fig:Figure3}c, closed circles). Upon decompression to the initial network height, the simulations reveal a strong increase of the shear modulus by a factor 140, while the density of nodes has increased by only a factor 1.32, suggesting that high levels of stiffening are possible with only a modest increase in cross-linking. The extent of stiffening observed in simulations is sensitive to the remodeling distance $d$: for a ten-fold decrease in $d$ ($d=0.001 l_0$) the stiffening factor is $\sim$10-fold, consistent with what we observe experimentally for fibrin networks. Importantly, this level of increase in stiffness is accompanied by only $\sim$3\% increase in cross-link density. We also find good qualitative agreement with experiment in the dependence of both the normal stress and shear modulus on compressive strain, as shown in Fig. \ref{fig:Figure3}d.

From the simulations, we can obtain insight into the mechanism of stiffening by considering the normal stress. During compression there is no significant change in the normal (axial) stress (Fig. \ref{fig:Figure3}c, open circles). However, upon decompression back to the original state, there is a progressive build-up of a negative (contractile) normal stress. We observe similar behavior in experiments, as shown in Fig. \ref{fig:Figure3}d. These observation are consistent with bond formation in the compressed state that results in a stressed configuration in the decompressed state. Importantly, the residual stresses in the decompressed state are expected to be tensile, corresponding to contractile forces acting on the initial uncompressed state. 

\begin{figure*}[ht]
	\begin{center}
		\centerline{\includegraphics[width=.7\textwidth]{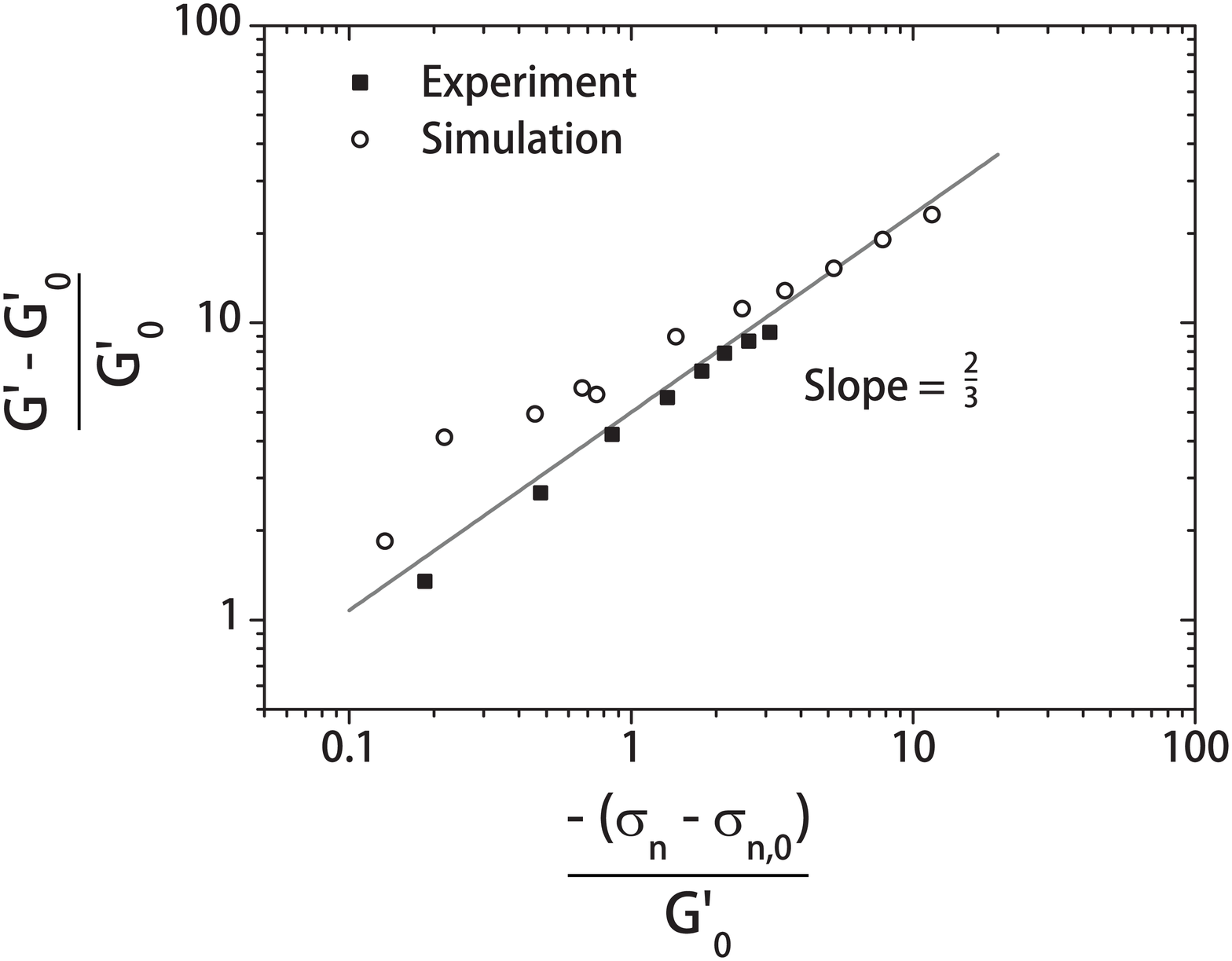}}
		\caption{\textbf{Universal scaling curve for the shear modulus of networks programmed by compression as a function of the built-in normal stress.} Increase in the shear modulus with normal stress for a 2 mg/ml fibrin gel (black triangles) and a simulated network with $\tilde{\kappa} = 10^{-4}$ (blue circles) trained by compressing to different levels of axial strain. The initial network parameters ($G'_0$, $\sigma_{n,0}$) are subtracted, and the data are normalized by dividing by $G'_0$. We find power-law scaling with a slope close to $2/3$ (black line, only to guide the eye).
		}
		\label{fig:Figure4}
	\end{center}
\end{figure*}

We can see that the degree of stiffening is directly related to the residual stress. In Fig. \ref{fig:Figure4}, we plot the increase in shear modulus against the normal stress observed for networks exposed to different levels of axial strain. We observe a power-law dependence of shear modulus on normal stress in the experiments and in the simulations. Importantly, this comparison between theory and experiment represents more than a similar scaling of stiffening with stress, since the scales for both stiffness and stress are identical. By normalizing both axes by the linear modulus, a direct and quantitative comparison between theory and experiment can be made. Interestingly, the observed power-law dependence is sublinear, with an exponent of $\sim$2/3. As noted above, the effect of compression suggests the presence of contractile stresses in the uncompressed state. A similar sublinear dependence of stiffening by random contractile forces was recently predicted theoretically for subisostatic networks, with an exponent of $\sim$0.8 \cite{Sheinman2012}. 

\section*{Conclusion}

The main finding of this Letter is that networks of adhesive fibers reinforce themselves under cyclic compressive loading by the spontaneous formation of new bonds between fibers. Compressive deformation is a particularly effective way to program networks compared to tensile and shear loading \cite{Kurniawan2016,Susilo2016,Cheema2007}, because compression causes densification and thereby increases the probability of fiber bonding. 
Our computational model shows that the rigidity of fibrous networks can be tuned over a wide range as a consequence of the inherent sensitivity of the stiffness of fibrous networks to variations in cross-link density and mechanical stress. A small increase in bond density of $\sim$3\% leads to a $\sim$10-fold increase in stiffness. Inelastic behavior thus provides a powerful way to program fibrous networks and tune their stiffness to diverse mechanical loading requirements. We expect this finding to be transferable to both tissue scaffolding proteins like collagen - which, similar to fibrin, can form bonds \cite{Nam2016}, and to cell scaffolding proteins like actin, which forms fiber bonds via dedicated cross-linking proteins \cite{Kasza2007}.  
Since our minimal model shows that the only ingredient needed to achieve programmable mechanics is fibers that are stiff and adhesive, this bioinspired design principle can be readily carried over to synthetic self-reinforcing materials using any of a wide range of available synthetic fibers, including carbon nanotubes and cellulose nanofibrils \cite{Davis2009,Olsson2010}.

\section*{Materials and methods}

\subsection*{Fibrin network rheology} We purchased chemicals from Sigma Aldrich (Zwijndrecht, Netherlands), human plasma fibrinogen and $\alpha$-thrombin from Enzyme Research Laboratories (Swansea, United Kingdom), and fibrinogen labeled with the fluorophore Alexa488 from Life Technologies (Eugene, OR, USA). Rheological experiments were performed using an Anton-Paar rheometer (Physica MCR501, Graz, Austria) equipped with a 40~mm stainless steel parallel plate geometry to allow variation of the gap size. Samples were prepared by diluting fibrinogen stock solution (dialysed against a 150mM NaCl, 20mM HEPES, pH 7.4 buffer) to 2 mg/ml in an assembly buffer to obtain final concentrations of 150mM NaCl, 20mM HEPES and 5mM CaCl$_2$ at a pH of 7.4. Polymerization was initiated by addition of 0.5 U/ml thrombin. Directly after mixing, we transferred the solution to the bottom plate of the rheometer and lowered the top plate to reach a gap size of 1.0~mm. During polymerization, the temperature was kept constant at 22\textdegree C, and solvent evaporation was prevented by adding a layer of mineral oil (M3516, Sigma Aldrich) on the liquid-air interface. 

During polymerization, we continuously measured the shear modulus, to verify that there was no evaporation or other disturbances of the sample. To this end, we applied a small oscillatory shear strain with an amplitude of 0.5\% and a frequency of 0.5~Hz, and measured the stress response. Polymerization was complete after 10 hours as indicated by a time-independent shear modulus. Compression was achieved by lowering the top plate at a slow rate of 1~$\mu$m/s in steps of 100~$\mu$m, corresponding to strain steps of 10\%. In between each compression step, we held the gap fixed for 125~s to allow the normal force to equilibrate (Supplementary Fig. \ref{fig:SIFigure6}). We then measured the shear modulus of the equilibrated network by applying a small oscillatory shear. Stepwise compression was continued until the thickness of the gel was reduced to 0.2~mm, corresponding to 20\% of the original height and an axial strain of 80\%. After compression, the gap was increased back to 1~mm (corresponding to an axial strain of zero) in a similar stepwise manner, and again allowing enough time for buffer to re-enter the network. We verified by UV-VIS absorption measurements of the expelled buffer solution that no protein was released from the networks during compression. In some experiments cross-linking by FXIII was inhibited by the addition of D004 (Zedira, Darmstadt, Germany) to the fibrinogen solution in a 30 to 1 molar ratio prior to the addition of thrombin \cite{Kurniawan2014}. For the rheology experiments, data points show the average of three measurements with error bars representing the standard deviation.

\subsection*{Optical tweezers} Optical tweezers experiments were performed using a custom-built four-trap optical tweezers system, combined with fluorescence microscopy. The sample chamber was a microfluidic flow cell with three inlets. Laminar flow conditions ensured that no mixing of the channels occurred. The first inlet contained $4.5\textrm{ \ensuremath{\mu}m}$ diameter spherical polystyrene beads (Spherotech, Lake Forest, IL, USA), which were trapped by a 20~W infrared ($\lambda=1064$~nm) laser. The second inlet contained fibrin fibers, obtained by adding 0.5~U/ml thrombin to a 0.01~mg/ml fibrinogen solution where 10\% of the monomers was labeled with Alexa-488. The third inlet contained just assembly buffer, which was supplemented in some experiments with the FXIII inhibitor D004. To test whether fibrin fibers are able to form bonds when brought in contact, we first trapped four beads in the bead channel, and next captured two fibrin fibers in the fibrin channel by connecting them end-to-end to a bead pair, using the ability of fibrin to strongly adhere to polystyrene beads \cite{Zeliszewska2014}. Finally, the two fibers were moved to the buffer channel using the beads as handles and one fiber was positioned underneath the other one by defocusing the trapping lasers and oriented in a crossed configuration. All beads were then brought back to the same focal plane in order to bring the fibers in contact. To test for fiber adhesion, we moved one of the bead pairs and performed time-lapse fluorescence imaging while concurrently measuring the force response on the other bead pair using a position-sensitive diode.

\subsection*{Simulations}
Disordered 2D lattices were generated by modifying triangular lattices with a lattice spacing of $l_0$ such that at each lattice vertex, one out of three passing filaments is randomly freed up \cite{Broedersz2011a}. This phantomization procedure sets the average connectivity (local coordination number) of the network to $z=4$. We further diluted the network by random removal of fiber segments until the average connectivity, $\langle z \rangle$, reached a value between 3 and 4 that is typical of biopolymer networks. Biopolymer networks of fibrin, collagen and actin usually exhibit combinations of fiber branching ($z=3$) and cross-linking ($z=4$) \cite{lindstrom2010biopolymer}. This average connectivity is below the point of marginal stability for a network of Hookean springs with only stretching interactions. This isostatic connectivity is $2d_s$ where $d_s$ is the dimensionality of the system\cite{maxwell1864calculation}. Spring networks are floppy below this threshold, but they can be stabilised by introducing a bending rigidity or by applying an internal or external stress \cite{wilhelm2003elasticity,Broedersz2011,Sheinman2012,Licup2015,alexander1998amorphous,DennisonPhysRevLett,Sharma2015a}. Here, we include bending interactions in the Hamiltonian of the system to account for the finite rigidity of fibrin fibers. Thus, the filaments in the network are described by the Hamiltonian
\begin{equation}
H=\sum_{f}\left[\int \frac{\kappa}{2} \left| \frac{d\hat{t}}{ds_{f}} \right|^{2} ds_{f} + \int \frac{\mu}{2} \left(\frac{dl}{ds_{f}} \right)^{2} ds_{f}\right]
\label{eqn:Efiber}
\end{equation}
where the first term accounts for bending and the second term accounts for stretching. The summation is over all filaments in the system, $\hat{t}$ is the unit tangent along the filament and $dl/ds_{f}$ is the longitudinal strain at point $s_f$ along the fiber contour. Each fiber is assigned a bending rigidity $\kappa$ and stretch modulus $\mu$. Note that the simulations assume an athermal network, which is appropriate for rigid biopolymers such as collagen, fibrin, microtubules, and actin bundles. We set the dimensionless fiber rigidity, defined as $\tilde{\kappa}=\kappa/(\mu l_0^2)$ where $l_0$ is the lattice spacing, to $10^{-4}$, a value that is appropriate for fibrin networks at the protein concentrations we are working at. The fiber rigidity is a dimensionless quantity that quantifies the relative importance of bend and stretch energy contributions and it is expected to vary approximately linearly with protein concentration assuming that the network architecture is invariant \cite{Licup2015,Sharma2015a}. The simulation results shown are an ensemble average of a total of ten different random networks.

\section*{Acknowledgements}
This work was supported by the Foundation for Fundamental Research on Matter (FOM), which is part of the Netherlands Organisation for Scientific Research (NWO), and by the National Science Foundation (Grant PHY-1427654).

\section*{Author contributions}
BEV, LCL, MV, GJL, EJG, NAK, FCM and GHK designed the research; BEV, AB and NAK conducted the experiments and analyzed data; LCL and MV developed the model and performed simulations. The authors jointly wrote the paper.

\section*{Competing financial interests}
The authors declare no competing financial interests.


\newpage

%
%
%
%

\newpage

\newpage

\newpage

\newpage

\newpage
\appendix

\newpage

\section*{Supplementary Information}

\section*{Estimation of fiber-fiber interaction strength by $\alpha$C-regions}

As explained in the main text, our findings indicate that fiber bond formation is mediated by noncovalent interactions. We hypothesize that the $\alpha$C-regions, two long and flexible chains which emanate from the distal ends of each fibrin monomer, are primarily responsible for this interaction, based on evidence from optical tweezers experiments showing strong interactions of these chains at the single molecule level \cite{Litvinov2007}. We can estimate the total strength $F$ of a bond between two adjacent fibers mediated by the two juxtaposed brushes of $\alpha$C-regions as: $F=P*f_{r}*\frac{d}{l_{m}}*\sqrt{n_{p}}*2$. Single-molecule force spectroscopy showed that two $\alpha$C-regions form bonds with a binding probability $P$ of 62\% and an average rupture force $f_{r}$ of 34~pN \cite{Litvinov2007}). The ratio $\frac{d}{l_{m}}$ is the fiber diameter $d\approx 100$~nm divided by the length of the fibrinogen monomer, $l_{m}=\textrm{45 nm}$ \cite{Fowler1981}, and gives the number of monomers over the length of the interaction area. We multiply $\frac{d}{l_{m}}$ by $\sqrt{n_{p}}$, where $n_{p}$ is the total number of protofibrils in a fiber cross section (around 65 \cite{Kurniawan2016}), to obtain an estimate of the total number of monomers per interaction area. Finally, the factor 2 takes into account that there are two $\alpha$C-regions per monomer.

This order-of-magnitude calculation predicts a binding strength of $760$~pN. Is this number large enough to make bond formation effectively irreversible even when fibrin networks are subject to a mechanical shear? To test this, we consider that for a 1\% deformation of a fibrin network with shear modulus 1700~Pa (1700~Pa being the average modulus after a compression-decompression cycle for our networks) we need to apply a 17~Pa shear stress. Per characteristic area of 2~$\mu$m $\times$ 2~$\mu$m (where 2~$\mu$m is an estimate of the pore size, or average area per fiber, based on \cite{Munster2013b}) we find that, to a first approximation, each fiber is subjected to an average force of 68~pN. Thus, we find that the newly formed connections are much stronger than the forces applied on the fibers, hence we can consider the new connections to be irreversible.

\newpage

\begin{suppfigure}[ht]
	\begin{center}
		\centerline{\includegraphics[width=.7\textwidth]{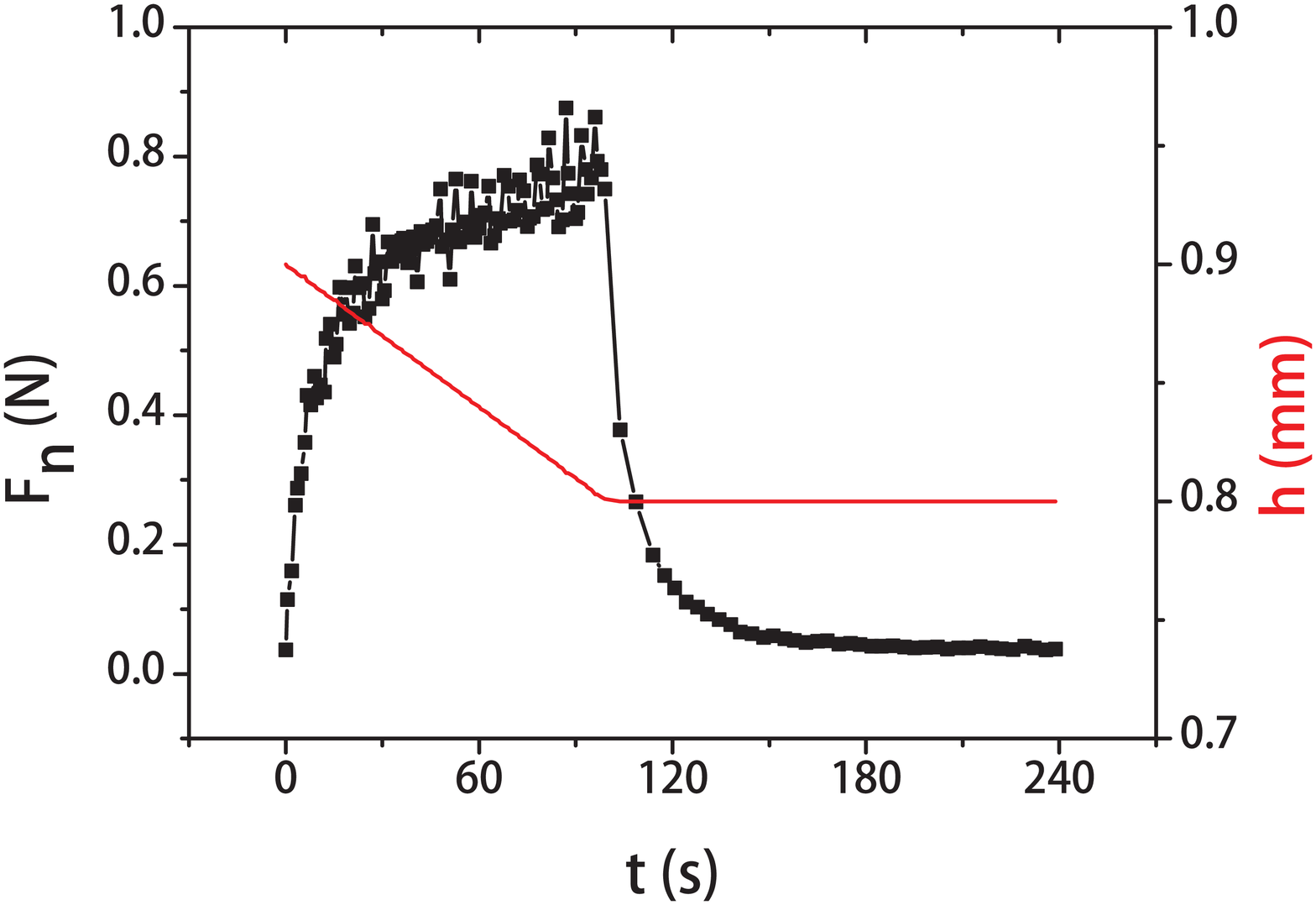}}
		\caption{\textbf{Normal force relaxation after a compressive step}. The normal (axial) force $F_n$, exerted by the fibrin network, equilibrates over time $t$, after the network underwent a stepwise compression from 0.9~mm to 0.8~mm. A small residual normal force (see Fig. 4 in the main text) corresponds to the build-up of internal normal stresses. The corresponding sample height $h$ is shown in red.
		}
		\label{fig:SIFigure6}
	\end{center}
\end{suppfigure}

\newpage

\begin{suppfigure}[ht]
	\begin{center}
		\centerline{\includegraphics[width=.7\textwidth]{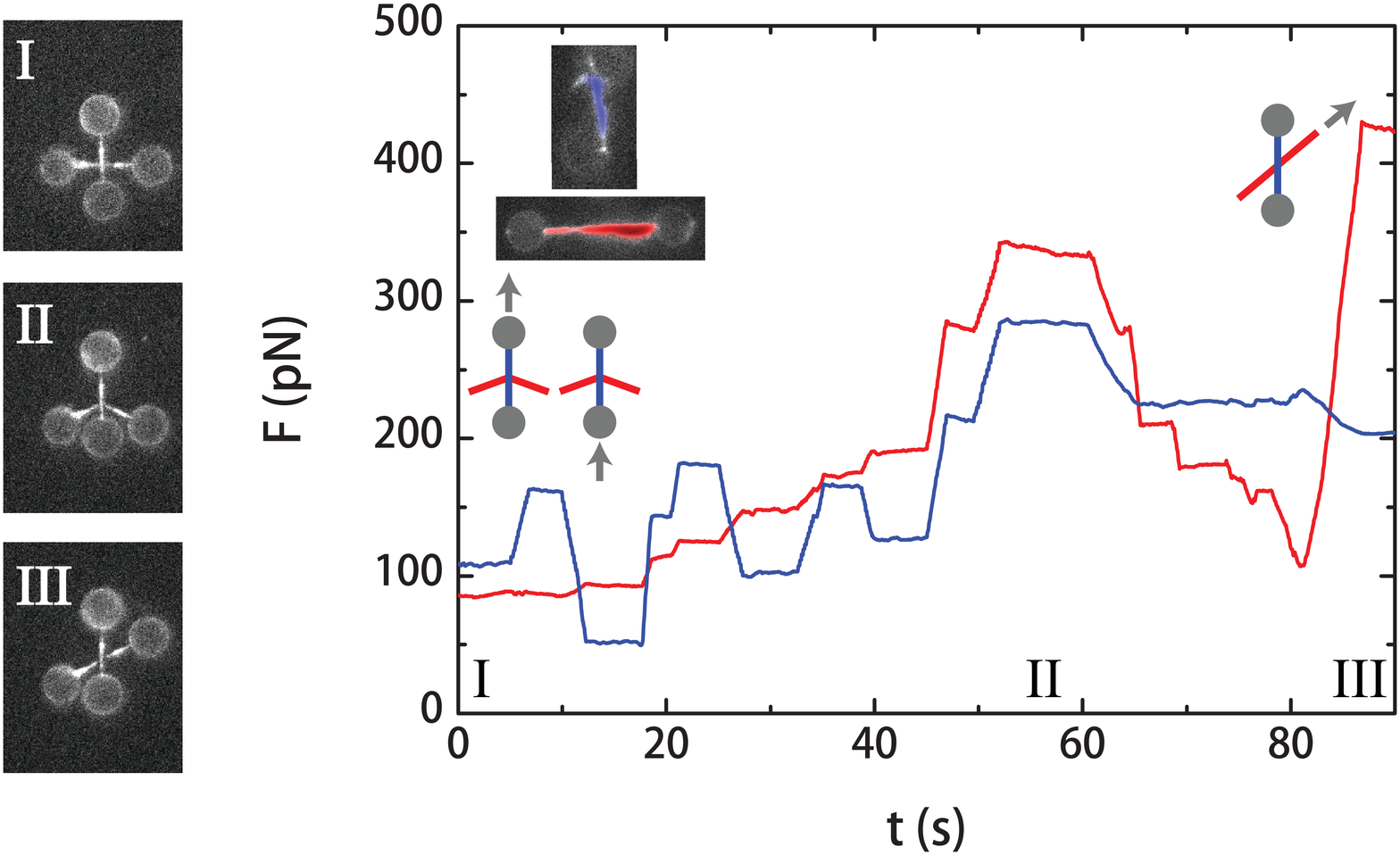}}
		\caption{\textbf{Optical tweezer measurement of the interaction between two fibrin fibers in the presence of D004, a specific inhibitor for the cross-linker FXIII}.  The numbers above the fluorescence images (left) correspond to time points indicated in the graph (right). The colours refer to the "horizontal" fiber (\textbf{red}) and the "vertical" fiber (\textbf{blue}). The fibers spontaneously form a strong (>300 pN) bond, indicating that bond formation does not require FXIII-mediated cross-linking.
		}
		\label{fig:SIFigure1}
	\end{center}
\end{suppfigure}

\newpage

\begin{suppfigure}[ht]
	\begin{center}
		\centerline{\includegraphics[width=.5\textwidth]{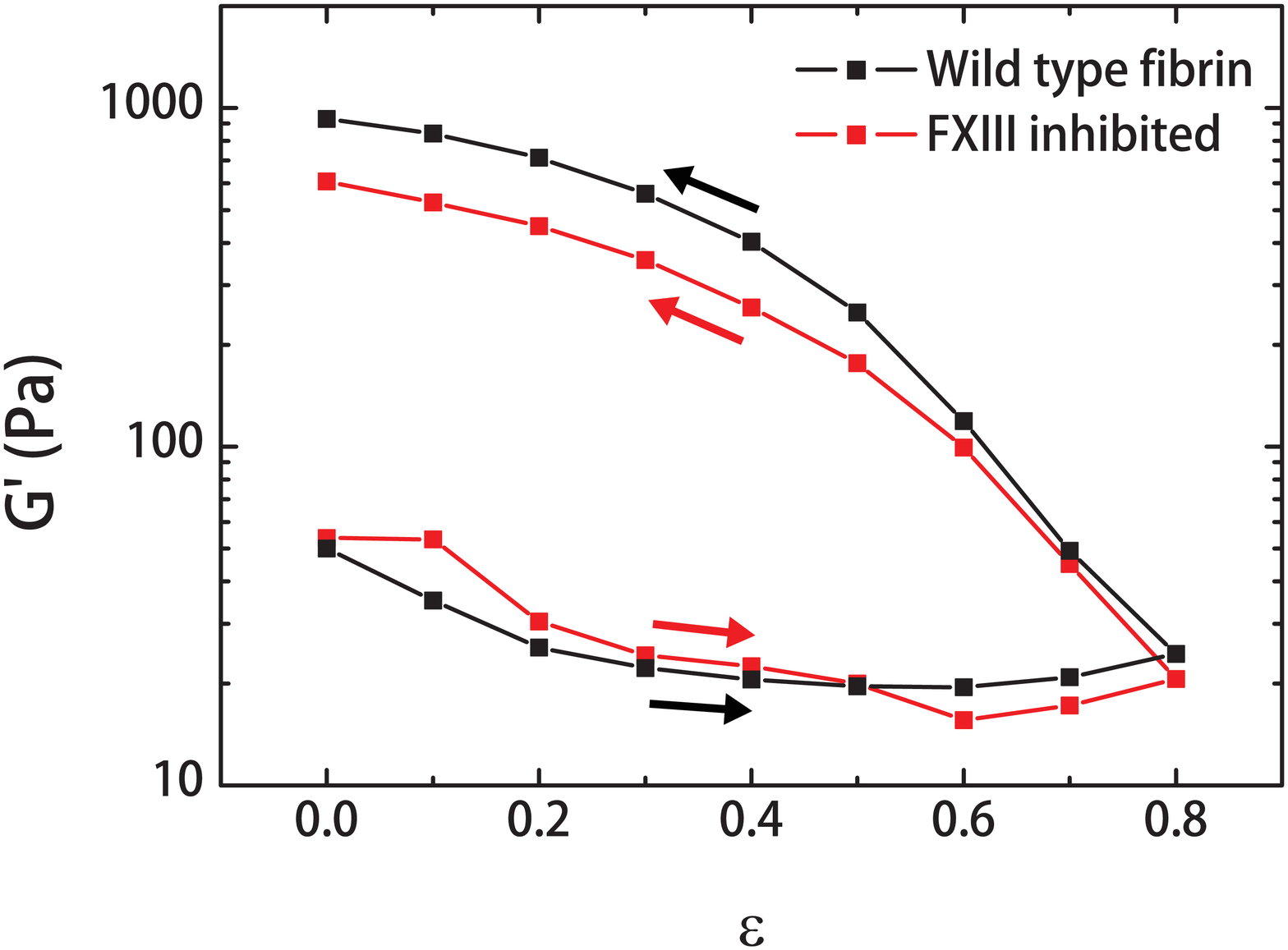}}
		\caption{\textbf{Compression and decompression of a 1 mg/ml fibrin gel, comparing cross-linked and uncross-linked networks.} The \textbf{black} line shows the response of a control network which s cross-linked by FXIII, while the \textbf{red} line shows the response of a corresponding gel where FXIII-mediated cross-linking is inhibited by adding D004. The initial sample height of the FXIII-inhibited gel was 0.5~mm; in order to match the compression rate to the other experiments, where the initial height was always 1.0~mm initial, the compressive speed was adjusted to 0.5~$\mu$m/s. Arrows indicate the sequence of the compression-decompression steps.
		}
		\label{fig:SIFigure2}
	\end{center}
\end{suppfigure}

\newpage

Supplementary Movie 1: \textbf{Direct measurement of the interaction between two individual fibrin fibers by optical tweezers.}
A movie showing the entire time-lapse sequence of an optical tweezers experiment. After capturing two fibers, one fiber is moved vertically and placed over the other fiber. The first fiber is lowered again to bring it in contact with the horizontally oriented fiber. We confirm that a bond is formed between the two fibers by moving any of the four beads, generating fiber bending and an increase in the trapping force. The diameter of the beads is 4.5 $\mu$m.

\bigskip

Supplementary Movie 1: \textbf{Two junctions merging in our computational model of adhesive fibrous networks.}
A section of a simulated network during compression, starting from an initial state with no applied axial strain to the moment where the first merging event occurs.

\end{document}